\documentclass[aps,prl,twocolumn,showpacs,superscriptaddress]{revtex4}
\usepackage{graphicx}
\usepackage{color}
\usepackage{amsmath,amssymb}

\begin{document}
\title{Crystalline and electronic structure of single-layer TaS$_2$}
\author{Charlotte E. Sanders}
\affiliation{Department of Physics and Astronomy, Interdisciplinary Nanoscience Center (iNANO), Aarhus University, 8000 Aarhus C, Denmark}
\author{Maciej Dendzik}
\affiliation{Department of Physics and Astronomy, Interdisciplinary Nanoscience Center (iNANO), Aarhus University, 8000 Aarhus C, Denmark}
\author{Arlette S. Ngankeu}
\affiliation{Department of Physics and Astronomy, Interdisciplinary Nanoscience Center (iNANO), Aarhus University, 8000 Aarhus C, Denmark}
\author{Andreas Eich}
\affiliation{Institute for Molecules and Materials, Radboud University,  6525 AJ Nijmegen, The Netherlands}
\author{Albert Bruix}
\affiliation{Department of Physics and Astronomy, Interdisciplinary Nanoscience Center (iNANO), Aarhus University, 8000 Aarhus C, Denmark}
\author{Marco Bianchi}
\affiliation{Department of Physics and Astronomy, Interdisciplinary Nanoscience Center (iNANO), Aarhus University, 8000 Aarhus C, Denmark}
\author{Jill A. Miwa}
\affiliation{Department of Physics and Astronomy, Interdisciplinary Nanoscience Center (iNANO), Aarhus University, 8000 Aarhus C, Denmark}
\author{Bj{\o}rk Hammer}
\affiliation{Department of Physics and Astronomy, Interdisciplinary Nanoscience Center (iNANO), Aarhus University, 8000 Aarhus C, Denmark}
\author{Alexander A. Khajetoorians}
\affiliation{Institute for Molecules and Materials, Radboud University,  6525 AJ Nijmegen, The Netherlands}
\author{Philip Hofmann}
\email{philip@phys.au.dk}
\affiliation{Department of Physics and Astronomy, Interdisciplinary Nanoscience Center (iNANO), Aarhus University, 8000 Aarhus C, Denmark}


\date{\today}
\begin{abstract}
Single-layer TaS$_2$ is epitaxially grown on Au(111) substrates.  The resulting two-dimensional crystals adopt the 1H polymorph.  The electronic structure is determined by angle-resolved photoemission spectroscopy and found to be in excellent agreement with density functional theory calculations. The single layer TaS$_2$ is found to be strongly \textit{n}-doped, with a carrier concentration of $0.3(1)$ extra electrons per unit cell.  No superconducting or charge density wave state is observed by scanning tunneling microscopy at temperatures down to 4.7~K.
\end{abstract}

\maketitle


Single layer (SL) transition metal dichalcogenides (TMDCs) share many fascinating properties with graphene. The electronic properties of the SL differ in subtle but important ways from those of the parent compounds \cite{Bollinger:2001aa,Mak:2010aa,Splendiani:2010aa,Novoselov:2005ab}. Most recent research on SL TMDCs has focused on semiconducting materials, because of the possibility to exploit spin and valley degrees of freedom \cite{Xu:2014ac}. 

Metallic SL TMDCs are interesting for other reasons: Their quasi-2D parent compounds host a wide range of symmetry-breaking electronic instabilities, such as charge density waves (CDWs), superconductivity (SC), and Mott states \cite{Wilson:1969aa,Rossnagel:2011aa}, and it is an open question how these would change in the SL limit. For CDWs driven  by electronic correlations or nesting, one might expect an increased transition temperature in the SL limit; but, since the CDW physics in the bulk is often complex, the opposite effect could occur, or an altogether different CDW periodicity might be found. This has been explored theoretically \cite{Calandra:2009aa,Ge:2012aa,Laverock:2013aa,Darancet:2014aa} and, very recently, experimentally for SL NbSe$_2$ \cite{Xi:2015,Ugeda:2016}. In SL NbSe$_2$, a strongly increased CDW transition temperature has been observed by optical techniques \cite{Xi:2015}, whereas atomically resolved scanning tunnelling microscopy (STM) measurements reveal a similar transition temperature as in the bulk  \cite{Ugeda:2016}. Such discrepancies might have several reasons. One is the possible role of contaminations in studies that are not performed in ultra-high vacuum (UHV). Another is the role of the substrate (silicon oxide in case of Ref. \cite{Xi:2015} and bilayer graphene in  Ref. \cite{Ugeda:2016}). For semiconducting TMDCs, the substrate can strongly modify the size of the SL TMDC's band gap via screening \cite{Ugeda:2014aa,Antonija-Grubisic-Cabo:2015aa}. This is probably less important for metallic TMDCs but the substrate can still give rise to doping and strain, two factors that significantly influence the formation of CDWs \cite{Friend:1987aa,Soumyanarayanan:2013aa}. 

A major challenge in the study of  thin TaS$_2$ and other metallic TMDCs is sample preparation.  In contrast with some other TMDCs, the material is air-sensitive---particularly in the atomically thin limit \cite{Navarro-Moratalla:2016}---and is therefore difficult to prepare by exfoliation.  Studies have been carried out on TaS$_2$ flakes exfoliated in air or in glove boxes \cite{Navarro-Moratalla:2016,CAO-Yu-Fei:2014aa,Galvis:2014aa}, and on flakes isolated by intercalation \cite{Ayari:2007aa}, with results that have been partly contradictory, in particular for the case of very thin films.

%

We overcome the reactivity issue by epitaxially growing SL TaS$_2$ on Au(111) under UHV conditions.  This allows for the preparation of atomically clean samples, along with a precise control of layer thickness.  Additionally, it results in a well-defined crystalline orientation in the TaS$_2$ with respect to the orientation of the underlying Au(111) substrate.  One question that arises is which of two possible structural phases---trigonal prismatic (hereafter referred to as ``1H'') or octahedral (``1T'')---will be adopted by SL TaS$_2$ (see insets in Fig. \ref{fig:3}).  This distinction is important, since the electronic properties and electronic instabilities  of the 2H and 1T bulk analogs are entirely distinct from one another \cite{Wilson:1969aa,Rossnagel:2011aa}. A critical question is whether the electronic instabilities observed in the bulk will also occur in the SL \cite{Darancet:2014aa}. We find that it is the 1H phase that is adopted in SL TaS$_2$  on Au(111) and that, surprisingly, neither CDW nor SC states are observable at temperatures as low as 4.7~K.

\begin{figure}
\includegraphics[width=0.5\textwidth]{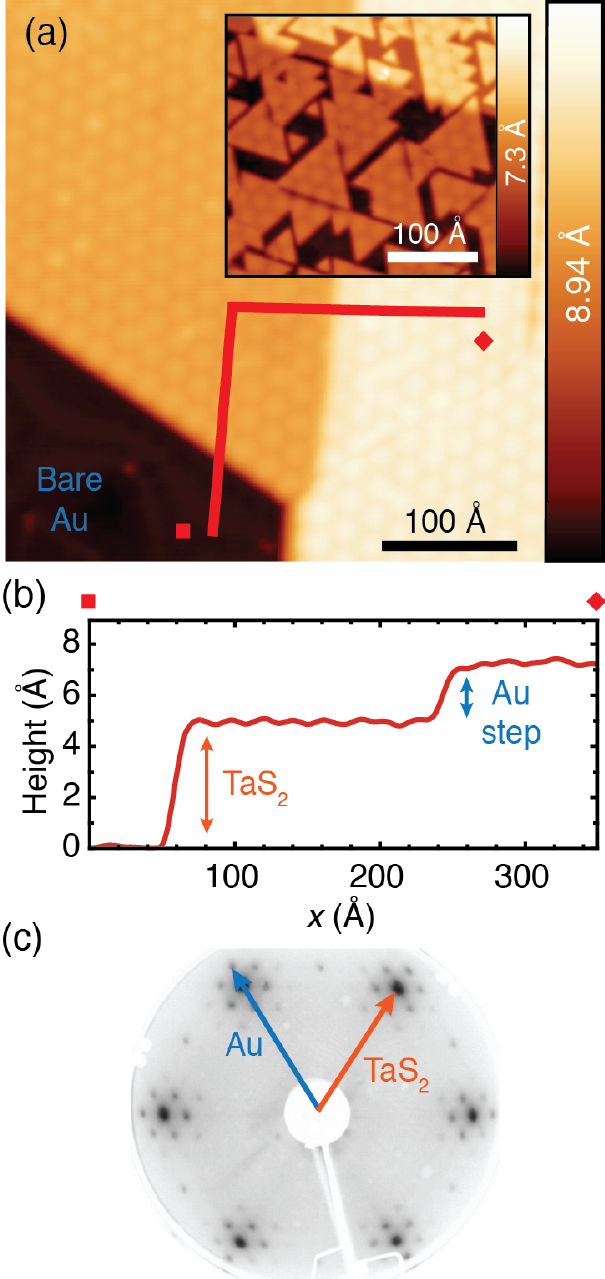}\\
\caption{(Color online) Morphology of SL TaS$_2$ islands grown on Au(111). (a)  STM scans showing sub-monolayer coverage for typical samples grown under different conditions to yield large areas (main panel) or smaller ordered islands (inset) of SL TaS$_2$. (Main panel:  $I_{set}$=105~pA, $V_S$=440~mV; inset:  $I_{set}$=199~pA, $V_S$=1.0~V.)  (b) Line profile taken along the line marked in panel (a).  Square and diamond symbols in (a) mark the start and end of the line.  The Au(111) step height and the apparent height of the TaS$_2$ SL are indicated.  (c) LEED data ($E_{kin}$=65.3~eV) from sample at sub-monolayer coverage.  Au and TaS$_2$ diffraction patterns are both visible, along with the pattern derived from the moir{\'e} superstructure.  Higher-order moir{\'e} spots can also be seen---in particular, halfway between the integer-order spots. } 
 \label{fig:1}
\end{figure}

The samples were synthesized using methods that are well-established for the growth of semiconducting SL TMDCs such as MoS$_2$ and WS$_2$; these methods are essentially based on the evaporation of a transition metal in an atmosphere of H$_2$S  onto a clean Au(111) surface that had been sputtered and annealed to exhibit the regular herringbone reconstruction \cite{Miwa:2015aa,Groenborg:2015,Dendzik:2015,Martinez:2016}.  Samples were grown and analyzed \textit{in situ} with angle-resolved photoemission spectroscopy (ARPES), low-energy electron diffraction (LEED), and STM at the SGM3 end-station of the ASTRID2 synchrotron radiation facility \cite{Hoffmann:2004aa}.  The sample temperature was 95~K for ARPES and LEED measurements. Low-temperature STM and scanning tunneling spectroscopy (STS) measurements were performed at 4.7~K in a separate chamber, to which the samples were transferred without breaking vacuum.

Density functional theory (DFT)  calculations for free-standing SL TaS$_2$ were performed using the VASP code \cite{Kresse:1993v1,Kresse:1996v2,Kresse:1996v3}. The valence electrons were described by plane-wave basis sets with a kinetic energy threshold of 415\,eV. The interaction between the valence and frozen core-electrons was accounted for by means of the projector augmented wave  method \cite{Blochl:1994}. The Perdew-Burke-Ernzerhof (PBE) approximation to the exchange-correlation functional was used \cite{Perdew:1996}. The optimized lattice parameter for the 1H and 1T phases of SL TaS$_2$ was found to be 3.337 and 3.372\,\AA, respectively. The TaS$_2$ SLs were modeled by single (1$\times$1) unit cells and the reciprocal space was sampled with a (20$\times$20$\times$1) mesh of $k$-points. Electron density was self-consistently converged with an energy threshold of $10^{-6}$\,eV. Atomic positions were relaxed until the forces on all atoms were smaller than 0.01\,eV\r{A}$^{-1}$. Spin-orbit coupling has been taken into account for all calculations.

\begin{figure}
\includegraphics[width=0.5\textwidth]{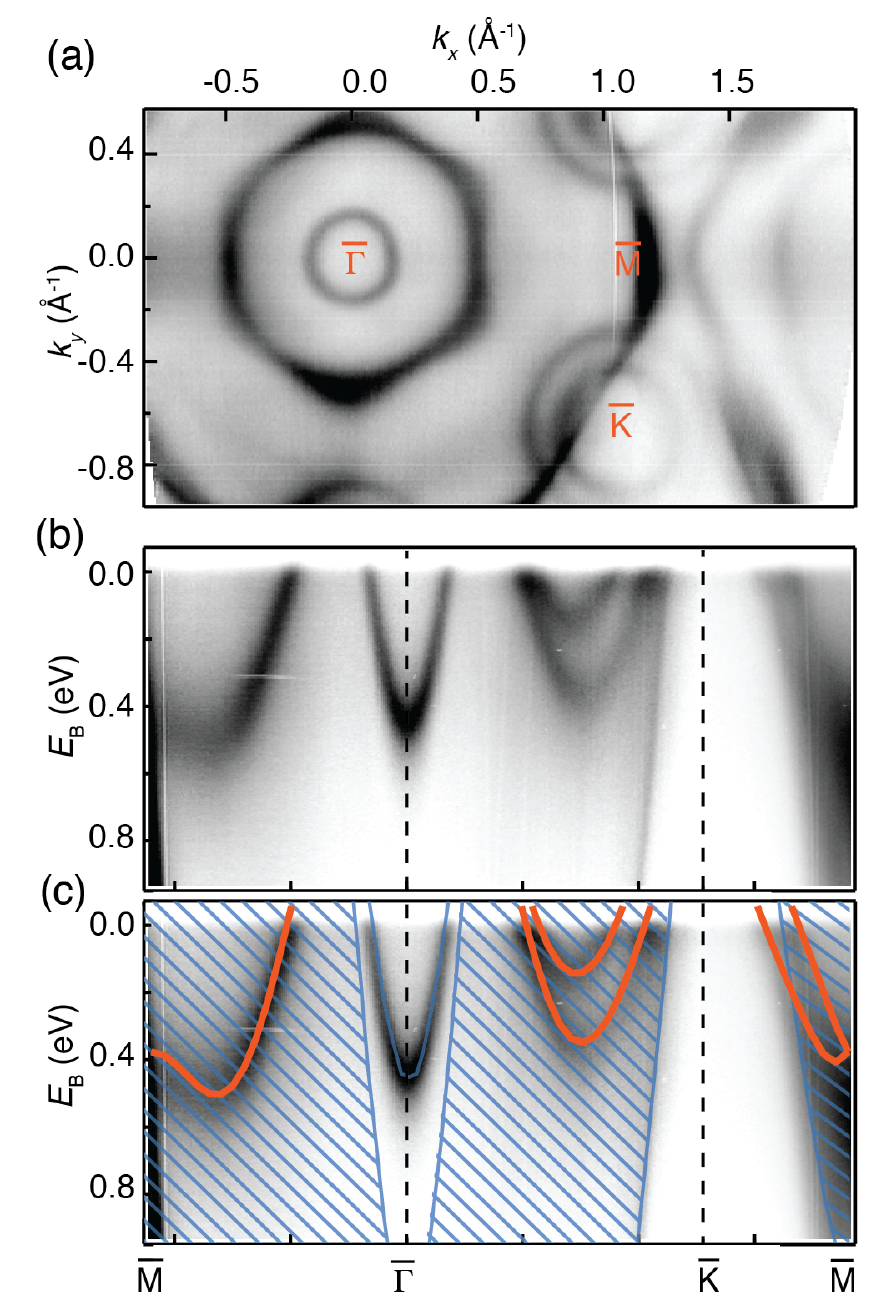}\\
\caption{(Color online) Electronic structure of SL TaS$_2$ measured with ARPES.  (a) Photoemission intensity at the Fermi energy (h$\nu$=30~eV).   (b) Photoemission intensity along high-symmetry directions of the 2D BZ (h$\nu$=30~eV).  (c) Data in (b) with the calculated 1H-TaS$_2$ band structure superimposed in orange. The calculated bands have been shifted by 0.12~eV to higher binding energy.  The Au surface state and projected bulk bands of Au(111) are indicated with blue, as guides to the eye \cite{Takeuchi:1991}.}
\label{fig:2}
\end{figure}

\begin{figure}
\includegraphics[width=0.5\textwidth]{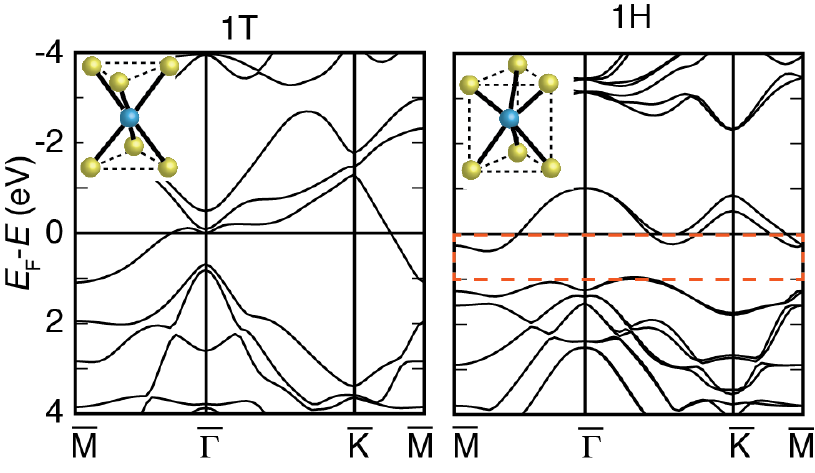}\\
\caption{(Color online) Calculated electronic band structures of SL 1T- and 1H-TaS$_2$. The atomic structures of the two phases are shown schematically in the insets. In the 1H band structure, the calculated spin-splitting at $\bar{K}$ is 0.348~eV. The orange dashed rectangle marks the energy and $k$-range studied in Fig. \ref{fig:2}(b) and (c).}
\label{fig:3}
\end{figure}

\begin{figure}
\includegraphics{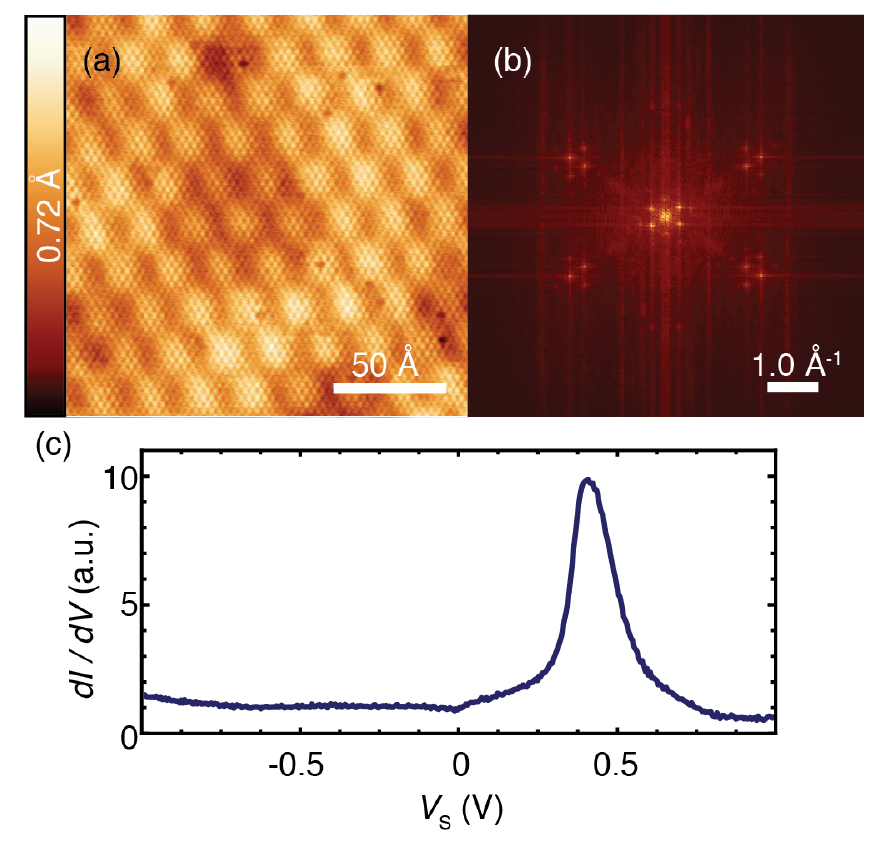}\\
\caption{(Color online) Absence of CDW state, revealed by STM/STS data acquired at 4.7~K.  (a) High-resolution STM image, showing atomic lattice and moir{\'e} superstructure.  ($I_{set}$=500~pA, $V_S$=4.3~mV.)  (b) Fast Fourier transform of data in panel (a).  (c) Representative STS point spectrum  ($I_{stab}$=500~pA, $V_S$=1.0~V, $V_{mod}$=5~mV, $f_{mod}$=4.423~kHz).}
\label{fig:4}
\end{figure}

Fig. \ref{fig:1}(a) shows the morphology of SL TaS$_2$ measured by STM for two samples of sub-monolayer coverage. For the main panel, the growth conditions have been chosen such as to produce large islands, whereas the inset shows conditions that give rise to many smaller and triangular clusters.  The large-scale hexagonal structure visible on the islands in Fig. \ref{fig:1}(a) is caused by a  moir{\'e} superstructure, similar to that observed for MoS$_2$/Au(111) \cite{Groenborg:2015} and WS$_2$/Au(111) \cite{Dendzik:2015},  with a superstructure lattice constant of $23.1(4)$~{\AA}. The moir{\'e}'s structure is more clearly seen in the atomically resolved STM image in Fig. \ref{fig:4}(a), and leads to satellite spots in the LEED image of Fig. \ref{fig:1}(c). LEED and STM results consistently give an in-plane atomic lattice constant of $3.3(1)$ {\AA} for the TaS$_2$ layer, which is the same as in the bulk compounds \cite{Wilson:1969aa}.

Note that it is difficult to distinguish between the 1H and 1T phases from such data, since both phases have hexagonal structures with similar lattice constants \cite{Qiao:2007}.  It is evident from the LEED pattern that the TaS$_2$ overlayer possesses a  well-defined orientation with respect to the underlying substrate. This orientation still permits the existence of two rotational domains (rotated by 180$^{\circ}$ with respect to each other). The up-pointing and down-pointing triangles in the inset to Fig. \ref{fig:1}(a) are distinguished by these two orientations \cite{Lauritsen:2007aa}. 

Although non-trivial interaction between the TaS$_2$ and the Au(111) is suggested by the well-defined crystalline orientation of the overlayer relative to the substrate, triangular islands are nevertheless observed to readily cross step edges, as in the inset to Fig. \ref{fig:1}(a).  This growth mode has been identified previously in graphene on certain substrates, such as Ir(111) \cite{Coraux:2008}, and has been interpreted as a consequence of weak substrate-overlayer interaction.

The ARPES data in Fig. \ref{fig:2} reveal the electronic band structure close to the Fermi level.   In addition to the features arising from the exposed Au substrate (in particular, the Au surface state at $\bar{\Gamma}$ and the Au \textit{sp} states closer to the edge of the Brillouin zone (BZ)), the TaS$_2$ overlayer exhibits a Fermi contour (Fig. \ref{fig:2}(a)) consisting of two distinct features. The first is an apparently  hexagonal  contour around the BZ centre $\bar{\Gamma}$. The second consists of two concentric rings around the $\bar{K}$ point.

The structure of the SL, 1H or 1T, can readily be determined by comparison to the DFT calculations for free-standing SL TaS$_2$ shown in Fig. \ref{fig:3}. While both the 1H and 1T structures give rise to a metallic SL, the actual band structures are very different. Clearly, the band structure for the 1H modification gives better  agreement with the experimental data because it contains the same Fermi contour features, while the Fermi contour for the 1T phase is very different \cite{SMAT}. Indeed, even quantitative agreement can be obtained when the calculated bands are shifted by $0.12(2)$~eV to higher binding energy to account for electron doping. This is shown in  Fig. \ref{fig:2}(c) (c.f. the region enclosed with a dashed orange box in Fig. \ref{fig:3}), where the shifted calculated bands are superimposed on the data. Thus, the structural phase preferred by epitaxial TaS$_2$ on Au(111) is 1H, rather than 1T.  As can be seen by inspection of the calculated 1H band structure in Fig. \ref{fig:3}, the features at the Fermi surface stem from the same band, which is spin-degenerate near $\bar{\Gamma}$ but strongly split near $\bar{K}$.  An inspection of the dispersion along different high-symmetry directions of the BZ (Figs. \ref{fig:2}(b) and \ref{fig:3}) shows that all Fermi contour features are hole pockets. A comparison to the calculated Fermi contour \cite{SMAT} reveals that the finite but unresolved splitting of the bands near the Fermi contour around $\bar{\Gamma}$ is responsible for the apparently hexagonal shape of this hole pocket, even though the individual bands do not have hexagonal Fermi contours. 

While the band structure of the bulk 2H parent compound can be considered to be quasi-2D, the truly 2D situation in the SL manifests important differences from the quasi-2D bulk case. Particularly relevant are the modifications to the single band forming the Fermi contour of the SL. In the SL, the band is two-fold degenerate near $\bar{\Gamma}$ and spin-split near $\bar{K}$. In the 2H bulk, on the other hand, the spin degeneracy is never lifted because of the structure's inversion symmetry. Still, the interaction of the two layers in the unit cell splits the  band into two two-fold degenerate bands near $\bar{\Gamma}$ while it remains four-fold degenerate at the BZ border point H. This causes a rather strong dispersion perpendicular to the TaS$_2$ layers, giving rise to a deviation from 2D behaviour   \cite{Barnett:2006aa}. Naively, one might thus expect a stronger tendency for the formation of CDW states in the SL, at least for nesting-driven CDWs.

We address the question of whether the sample exhibits CDW or SC by inspecting STM/STS data taken at 4.7~K.  In bulk 2H-TaS$_2$, the superconducting critical temperature is $T_C$= 600~mK \cite{Garoche:1978}, and a CDW of $(3 \times 3)$  periodicity sets in below $T_{CDW}$=75~K \cite{Wilson:1969aa}, with an accompanying lattice distortion that is clearly visible as a periodic superstructure in the STM data \cite{Guillamon:2011aa} (the same is true for SL NbSe$_2$ \cite{Ugeda:2016}).  The low-temperature STM data in Fig. \ref{fig:4}(a), on the other hand, show no indication of any additional periodicities apart from the lattice as such and the moir{\'e} superstructure. This is confirmed by an inspection of the Fourier transformation of  the image, which only shows these two periodicities (see Fig. \ref{fig:4}(b)). STS measurements made at 4.7 K show a strong feature at approximately 430~meV above the Fermi energy, consistent with results obtained from NbSe$_2$, where this  has been associated with the top of the valence band at $\bar{\Gamma}$ \cite{Ugeda:2016}). The spectra give no indications of a SC gap. 

It is not surprising that SC is not observed, since the $T_C$=600~mK is much lower than the measurement temperatures used in the present study.  One set of recent studies has suggested that $T_C$ in thin flakes might be higher than the bulk value \cite{Galvis:2014aa,Navarro-Moratalla:2016}---in contrast to what has been seen for the case of NbSe$_2$, where $T_C$ is  suppressed in the SL limit \cite{Xi:2015,Ugeda:2016}---but even the highest proposed temperatures for  thin TaS$_2$ are considerably smaller than the measurement temperatures of the present study.

The absence of a CDW, however, is surprising.  The onset of CDW instability at  75~K in bulk 2H-TaS$_2$, though it is below the temperature at which ARPES and LEED data were collected here, is significantly above that at which STM and STS data were acquired.  In the related material NbSe$_2$, there have been conflicting findings for the CDW onset temperature in the SL with respect to the bulk. A strongly increased $T_{CDW}$ was reported for SL NbSe$_2$ on silicon oxide \cite{Xi:2015}, whereas a minor decrease  was observed for SL NbSe$_2$ in UHV on bilayer graphene \cite{Ugeda:2016}.)

In this context, it is interesting to compare the Fermi vector $2 k_F$=0.96(2)~\AA$^{-1}$ measured across the hole pocket at $\bar{\Gamma}$ in the present study with that which would be required if a $(3 \times 3)$ CDW state were driven  by nesting:  in the nesting-driven case, the nesting vector would need to be 0.73~\AA$^{-1}$.  Clearly, this value matches poorly to the experimentally derived value of $2 k_F$; however, this disagreement is not sufficient to explain the absence of a CDW, since simple nesting cannot explain the CDW in the bulk parent materials, either \cite{Rice:1975,Castro-Neto:2001aa,Rossnagel:2001aa,Calandra:2009aa}. 

The most likely explanation for the lack of CDWs is doping of the TaS$_2$ by the Au substrate. As already seen in Fig. \ref{fig:2}(c), the calculated bands have to be shifted to higher energy by 0.12~eV to match the observed dispersion, suggesting that the SL is electron-doped. A rigorous determination of the Fermi contour areas gives a carrier concentration of approximately $0.3(1)$ extra electrons per unit cell \cite{SMAT}---i.e., an occupation of $1.3(1)$ electrons in the uppermost valence band, in contrast with the single electron that one would expect for the undoped material.  Previous studies of alkali-intercalation compounds \cite{Friend:1987aa} have shown that the CDW can already be suppressed at more modest electron doping, suggesting that this plays a decisive role. 

The CDW transition might also be influenced by other factors:  e.g., reduced dimensionality; substrate interactions other than doping, such as screening \cite{Ugeda:2014aa,Antonija-Grubisic-Cabo:2015aa}; chemical bonding \cite{Dendzik:2015}; or strain \cite{Soumyanarayanan:2013aa}. In the present case, the uncertainty the measurement of the atomic lattice puts an upper limit of $\approx$3\% on the in-plane strain and Figs. \ref{fig:2} and \ref{fig:3} show that the substrate has only a minor influence on the band structure of SL TaS$_2$, apart from the doping. However, these factors might still play a minor role in suppressing CDW formation \cite{Ge:2012aa}.  

In summary, we have successfully used an epitaxial approach to fabricate monolayer TaS$_2$ on Au(111) substrates.  We have investigated band structure and crystallinity \textit{in situ} using ARPES, STM and LEED.  We have used low-temperature STM/STS to obtain detailed information on the growth mode and to measure the density of states close to the Fermi level.  Comparing our band structure measurements to calculations by DFT, we have determined that our samples are in the 1H phase.  We do not see evidence of SC or a CDW state at temperatures down to 4.7~K. It remains, of course, possible that CDW or SC transitions are observed at lower temperatures.


We gratefully acknowledge financial support from the VILLUM foundation, the Danish Council for Independent Research, Natural Sciences under the Sapere Aude program (Grant Nos. DFF-4002-00029, and 0602-02566B), the Lundbeck Foundation. AB acknowledges support from the European Research Council under the European Union's Seventh Framework Programme (FP/2007-2013) / Marie Curie Actions / Grant no. 626764 (Nano-DeSign). AE and AAK acknowledge financial support from the Emmy Noether
Program (KH324/1-1) via the Deutsche Forschungsgemeinschaft, and from FOM which is part of NWO, and the NWO Vidi program. 



\end{document}